\documentclass[reprint,prb,aps,amsmath,amssymb,superscriptaddress,floatfix,longbibliography]{revtex4-2}

\usepackage{graphicx}
\usepackage{bm}
\usepackage[colorlinks=true,linkcolor=blue,citecolor=blue,urlcolor=blue]{hyperref}
\usepackage{xurl}
\usepackage{tikz}
\usetikzlibrary{arrows.meta,decorations.pathmorphing,calc}
\usepackage{xcolor}
\usepackage{soul}
\soulregister{\cite}{7}
\soulregister{\ref}{7}
\soulregister{\eqref}{7}

\newcommand{\Aedge}{A_{\mathrm{edge}}}
\newcommand{\Cbar}{\overline{C}_{0\to j}}
\newcommand{\thalf}{\tau_{1/2}}
\newcommand{\tedge}{\tau_{\mathrm{edge}}}

\begin{document}

\title{Bottlenecked scrambling of boundary information in nonintegrable Ising chains}

\author{Rohit Kumar Shukla}
\email[]{rohitkrshukla.rs.phy17@itbhu.ac.in}
\affiliation{Department of Chemistry, Bar-Ilan University, Ramat-Gan, 5290002, Israel}
\affiliation{Research Institute CODE, University of the Bundeswehr Munich, D-81739 Munich, Germany}

\date{\today}

\begin{abstract}
Almost strong zero modes protect the memory of boundary spins in Ising chains, even at infinite temperature. However, how this long-lived boundary information is transferred to and scrambled by the bulk remains poorly understood. Here, we investigate this process using infinite-temperature out-of-time-order correlators (OTOCs) in a nonintegrable transverse-field Ising chain with next-nearest-neighbor interactions. Combining exact diagonalization and dynamical typicality, we find that edge-to-bulk scrambling is governed primarily by the slow decay of boundary memory rather than by information propagation through the bulk. This behavior is captured by a parameter-free time-doubling relation between the bulk-averaged edge-to-bulk OTOC and the edge autocorrelation function, $\Cbar(t)\simeq1-\Aedge(2t)$. Consequently, the loss of boundary memory directly determines the timescale for scrambling across the chain. We further show that this bottleneck depends on the operator, with operators that overlap with the almost strong zero mode exhibiting slow scrambling, while those without such overlap scramble on bulk dynamical timescales. Finally, we find that the edge lifetime depends sensitively on temperature and the sign of the next-nearest-neighbor interaction, with cooling enhancing or suppressing boundary protection depending on whether the interaction is ferromagnetic or frustrating.
\end{abstract}

\maketitle

\section{Introduction}
\label{sec:intro}

Quantum information placed into an interacting many-body system, for example by flipping a single spin, is rapidly converted into nonlocal correlations that no local measurement can recover. This process, \emph{scrambling}, connects quantum chaos, thermalization of isolated systems, and the information problem of black holes~\cite{shenker2014,roberts2015,maldacena2016bound,sekino2008,hayden2007,hosur2016,swingle2018,xu2024tutorial}. Its standard diagnostic is the out-of-time-order correlator (OTOC)~\cite{shenker2014,maldacena2016bound,larkin1969}. In lattice systems with short-range interactions, information spreads within a light cone set by the butterfly velocity, bounded by the Lieb--Robinson velocity~\cite{roberts2015,lieb1972}, and random-circuit models and hydrodynamic arguments give a largely universal description of this front~\cite{aleiner2016,nahum2018,vonkeyserlingk2018,khemani2018,rakovszky2018,xu2020mpo}. OTOCs have been studied in Ising~\cite{lin2018ising,craps2020,heyl2018}, diffusive and long-range~\cite{bohrdt2017,luitz2017,pappalardi2018}, random-field and Floquet~\cite{riddell2019,shukla2020epl,shukla2022,shukla2022pra,shukla2025epjb}, and integrable or chaotic chains~\cite{huang2019finite,garciamata2018,fortes2019}. Scrambling is slowed down in many-body-localized~\cite{fan2017,huang2017mbl,swingle2017slow,lee2019} and spatially inhomogeneous~\cite{malik2025} systems. The late-time OTOC is tied to eigenstate thermalization~\cite{deutsch1991,srednicki1994,rigol2008,dalessio2016}, operator growth can be organized in Krylov space~\cite{parker2019}, and scrambling has been measured on several experimental platforms~\cite{li2017,garttner2017,landsman2019,joshi2020,mi2021,braumuller2022,swingle2016measuring}.

Boundaries can behave very differently from the bulk. The transverse-field Ising chain is equivalent to Kitaev's Majorana wire~\cite{pfeuty1970,kitaev2001,alicea2012,nayak2008}, and its ordered phase hosts a \emph{strong zero mode}, an edge-localized operator that commutes with the Hamiltonian~\cite{fendley2012,fendley2016}, so that an edge spin remembers its orientation even at infinite temperature~\cite{kemp2017}. When integrability is broken, the memory becomes finite but remains very long. Kemp \emph{et al.}\ traced this to an \emph{almost} strong zero mode~\cite{kemp2017}, which Else \emph{et al.}\ identified as a prethermal phenomenon with a nearly exponential lifetime~\cite{else2017}, in line with general results on prethermalization~\cite{abanin2017,mori2018,bertini2015,mallayya2019}. Yates, Abanov, and Mitra computed the lifetimes in a Krylov-space picture, in which the edge operator is a quasibound state that slowly tunnels into a delocalized, ``metallic'' bulk~\cite{yates2020,yates2020dyn,yates2021krylov}; the same physics governs Floquet edge modes~\cite{yates2019,yates2022} and boundary impurities~\cite{yeh2023}. The underlying mechanism is an energy mismatch: flipping the edge spin changes the number of domain walls by one, which the bulk can absorb only in a high-order process~\cite{kemp2017,yates2020,kemp2020}. For the Ising chain with a ferromagnetic next-nearest-neighbor interaction studied here, Kemp, Yao, and Laumann showed that the edge-memory lifetime can become parametrically large when the transverse field and next-nearest-neighbor interaction are weak compared with the dominant nearest-neighbor coupling~\cite{kemp2020}. Related physics appears in localization-protected edge order~\cite{huse2013,bahri2015} and interface zero modes~\cite{olund2025}, finite temperatures have been studied with tensor networks~\cite{tausendpfund2025}, and prethermal edge modes have been observed on a superconducting processor~\cite{jin2025}.

OTOCs near boundaries have been investigated in several classes of systems, including Majorana-type chains, where edge-to-edge OTOCs exhibit a ``prescrambling'' plateau~\cite{dag2020}, free-fermion~\cite{sedlmayr2023} and driven $XY$~\cite{sur2024} chains, where information remains localized near the edge, and $\mathbb{Z}_3$ parafermion chains, where the OTOC exhibits a pronounced peak at the boundary~\cite{zhang2021parafermion}. These studies have established that boundary modes can protect edge operators against scrambling. However, {\it how the information protected at the boundary of a nonintegrable system is eventually released into the bulk remains poorly understood.} In particular, the space--time structure of edge-to-bulk scrambling and its relation to the edge-spin autocorrelation have yet to be explored. In bulk systems, OTOCs involving conserved densities are known to exhibit slow relaxation and closely follow the corresponding two-point correlation functions~\cite{khemani2018,rakovszky2018,balachandran2021,balachandran2023}. Since an almost strong zero mode acts as a quasiconserved quantity localized at the boundary, it is natural to ask whether a similar connection holds for edge OTOCs and autocorrelations.

\begin{figure}[t]
\centering
\begin{tikzpicture}[x=0.78cm,y=0.78cm, every node/.style={font=\footnotesize}]
  \fill[blue!7, rounded corners=3pt] (1.45,-0.62) rectangle (8.5,0.62);
  \node[blue!45!black] at (5.0,0.95) {chaotic bulk};
  \foreach \x in {1,...,8} \shade[ball color=blue!55!white] (\x,0) circle (0.15);
  \shade[ball color=red!75!white] (0,0) circle (0.2);
  \foreach \x in {0,...,7} \draw[line width=1.2pt, black!70] (\x+0.2,0) -- (\x+0.8,0);
  \foreach \x in {0,...,6} \draw[line width=0.5pt, black!45] (\x+0.12,0.17) .. controls (\x+1,0.48) .. (\x+1.88,0.17);
  \draw[very thick, red!60!black] (0.72,-0.38) -- (0.72,0.38);
  \draw[very thick, red!60!black] (0.9,-0.38) -- (0.9,0.38);
  \draw[-{Stealth[length=4pt]}, thick, red!60!black, decorate, decoration={snake,amplitude=0.8pt,segment length=4pt}] (0.3,-0.25) -- (1.6,-0.25);
  \node[red!55!black] at (0,-0.75) {$Z_0$};
  \node[black!60] at (4.5,-0.95) {$J$ (thick), $K$ (thin)};
\end{tikzpicture}
\caption{\textbf{Bottlenecked scrambling.} Schematic of the open chain of Eq.~\eqref{eq:H}. The boundary spin $Z_0$ (red) and the bulk spins (blue) are coupled by nearest-neighbor ($J$, thick lines) and next-nearest-neighbor ($K$, thin arcs) Ising interactions. The double bar marks the bottleneck between the edge and the bulk, and the wavy arrow indicates the slow leakage of edge information into the bulk (shaded), where it is scrambled quickly and uniformly.}
\label{fig:cartoon}
\end{figure}
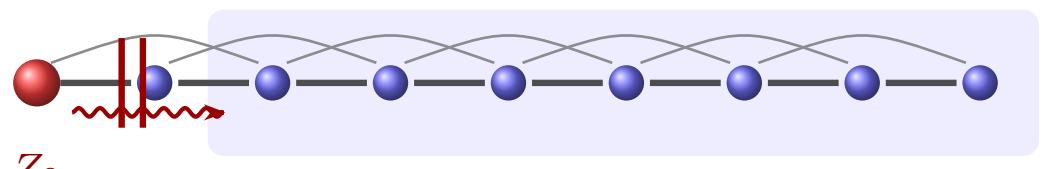

Here, we address these questions in the nonintegrable spin-chain model studied in Ref.~\cite{kemp2020}. We investigate how edge-localized information spreads into the bulk and how this process is reflected in the edge-to-bulk OTOC. We find that the dynamics is characterized by a bottleneck, with only a weak ballistic front propagating through the system. Beyond this front, the OTOC grows slowly and nearly uniformly across the bulk, indicating that its growth is governed primarily by the release of information from the boundary. We further establish a parameter-free time-doubling relation, $\Cbar(t)\simeq1-\Aedge(2t)$, between the bulk-averaged OTOC and the edge-spin autocorrelation, which follows from time-reversal symmetry and the separation between slow edge leakage and fast bulk scrambling. This relation provides a direct connection between bulk information spreading and the persistence of the edge-localized component of the evolving operator. We then identify the protected edge operators by constructing a zero mode through spectral filtering. Finally, we examine the finite-size dependence of the edge lifetime for system sizes up to $L=18$ and investigate how finite temperature and the sign of the next-nearest-neighbor coupling affect the protection.

The paper is organized as follows. Section~\ref{sec:model} introduces the model and observables. Section~\ref{sec:results} presents the results: bulk chaos and the long edge memory due to an almost strong zero mode (Secs.~\ref{sec:results}\,A--B), the space--time and operator structure of edge scrambling (Secs.~\ref{sec:results}\,C--D), the time-doubling law (Sec.~\ref{sec:results}\,E), and the lifetime of the slow component and its temperature dependence (Secs.~\ref{sec:results}\,F--G). Section~\ref{sec:conclusions} gives our conclusions and outlook, and the appendices contain the derivation and numerical details.
\section{Model and observables}
\label{sec:model}

\subsection{Model}

We consider an open chain of $L$ spin-$\frac12$ degrees of freedom, labelled by $j=0,\ldots,L-1$, governed by the Hamiltonian~\cite{kemp2020}
\begin{equation}
H=-J\sum_{j=0}^{L-2}Z_jZ_{j+1}+K\sum_{j=0}^{L-3}Z_jZ_{j+2}-g\sum_{j=0}^{L-1}X_j,
\label{eq:H}
\end{equation}
where $X_j$ and $Z_j$ are Pauli operators acting on site $j$. Here $J>0$ is the ferromagnetic nearest-neighbor Ising coupling, $K$ is the next-nearest-neighbor Ising coupling, and $g$ is the transverse magnetic field, which acts on all sites. A schematic illustration of the chain and its couplings is shown in Fig.~\ref{fig:cartoon}. We set $J=1$, thereby measuring energies in units of $J$ and times in units of $J^{-1}$.

For $K=0$, the model reduces to the integrable transverse-field Ising chain~\cite{pfeuty1970}, which is equivalent to a free Majorana wire~\cite{kitaev2001,alicea2012}. In its ordered phase, $g<1$, the chain supports strong zero modes localized at its two boundaries, which commute with $H$ up to corrections exponentially small in $L$~\cite{kitaev2001,fendley2012,kemp2017}. The left-edge mode is
\begin{equation}
\Psi_0=\sqrt{1-g^2}\,\sum_{k=0}^{L-1}g^{k}\,X_0X_1\cdots X_{k-1}Z_k ,
\label{eq:Psi0}
\end{equation}
where the $k=0$ term is $Z_0$ itself. Each further term extends the operator by one site and is suppressed by a factor $g$, so that $\Psi_0$ is exponentially localized at the left boundary. The prefactor normalizes $\Psi_0$, and its overlap with the edge spin is $|\mathrm{Tr}(Z_0\Psi_0)|^2/D^2=1-g^2$, where $D=2^L$ is the Hilbert-space dimension.

The next-nearest-neighbor interaction $K$, as in the axial next-nearest-neighbor Ising model~\cite{selke1988}, breaks integrability. Unless stated otherwise, we take $K=-0.3$, for which both Ising interactions favor ferromagnetic order. The Hamiltonian conserves the global spin-flip parity $P=\prod_j X_j$ and is invariant under spatial reflection $j\to L-1-j$. It is also real in the $Z$ basis, so complex conjugation in this basis implements time reversal.

In the ordered phase, the boundary spin $Z_0$ retains a substantial overlap with an almost strong zero mode $\hat\Psi$ [constructed explicitly in Eq.~\eqref{eq:Psi}] localized near the edge~\cite{fendley2016,kemp2017,else2017,yates2020}. The reason this mode is long-lived is energetic. Flipping the edge spin $Z_0$, when it is aligned with its neighbor $Z_1$, costs the Ising energy $2(J+|K|)$ if $Z_2$ is parallel to $Z_0$ and $2(J-|K|)$ if it is antiparallel~\cite{kemp2017,kemp2020}. For $g\ll2(J-|K|)$, a single flip induced by the transverse field is therefore off-resonant, and transferring the energy and information of the edge spin to the bulk requires high-order processes. This creates a bottleneck between the edge and the bulk and results in a long-lived boundary memory whose lifetime increases rapidly as $g$ decreases~\cite{parker2019edge,kemp2017,else2017,kemp2020}. Once the information escapes the edge, it is rapidly scrambled by the nonintegrable bulk~\cite{nahum2018,vonkeyserlingk2018,lin2018ising,craps2020}. Edge-to-bulk scrambling is therefore controlled primarily by the rate of information release from the boundary, rather than by its subsequent propagation through the chain (Fig.~\ref{fig:cartoon}).

An exact symmetry relates the dynamics at opposite signs of the next-nearest-neighbor coupling. Consider the unitary transformation
\begin{equation}
U=\Big(\prod_{j=0}^{L-1}Z_j\Big)\Big(\prod_{j\,\mathrm{odd}}X_j\Big).
\label{eq:U}
\end{equation}
The first factor reverses the sign of every $X_j$, and the second reverses $Z_j$ on the odd sites. Hence the nearest-neighbor terms $Z_jZ_{j+1}$ change sign, whereas the next-nearest-neighbor terms $Z_jZ_{j+2}$ do not, so that $UH(g,K)U^\dagger=-H(g,-K)$. Because site $0$ is even, the edge operator is left invariant, $UZ_0U^\dagger=Z_0$. As a result, all infinite-temperature dynamical quantities constructed from $Z_0$, including the OTOCs studied below, are identical for $K$ and $-K$. At finite temperature, the same transformation maps the dynamics of the frustrated chain ($K>0$, where the two Ising couplings compete) at inverse temperature $\beta$ onto that of the unfrustrated chain ($K<0$) at inverse temperature $-\beta$. Thus, the sign of $K$, which is immaterial for the infinite-temperature dynamics considered here, becomes important at finite temperature.

\subsection{Observables}

We denote the Heisenberg evolution of an operator $W$ by $W(t)=e^{iHt}We^{-iHt}$. All quantities are evaluated at infinite temperature, where the expectation value of an operator $O$ is $\langle O\rangle=\mathrm{Tr}(O)/D$. The \emph{memory} of an operator $W$ normalized as $\mathrm{Tr}(W^\dagger W)/D=1$ is quantified by its infinite-temperature autocorrelation function~\cite{kemp2017,yates2020},
\begin{equation}
A_W(t)=\frac{1}{D}\,\mathrm{Tr}\big[W^\dagger W(t)\big],\qquad \Aedge\equiv A_{Z_0}(t).
\label{eq:A}
\end{equation}
To characterize the spreading of quantum information, we use the out-of-time-order correlator (OTOC), defined through the squared commutator between a source operator $W$ and a probe operator $V$~\cite{maldacena2016bound,swingle2018,xu2024tutorial,larkin1969},
\begin{equation}
C_{W\to V}(t)=\frac{1}{2D}\,\mathrm{Tr}\Big([W(t),V]^\dagger\,[W(t),V]\Big).
\label{eq:C}
\end{equation}
For Hermitian unitary operators $W$ and $V$, this definition reduces to
\begin{equation}
C_{W\to V}(t)=1-\frac{1}{D}\,\mathrm{Re}\,\mathrm{Tr}\big[W(t)VW(t)V\big].
\label{eq:Cunitary}
\end{equation}
The commutator definition is more general and also accommodates the nonunitary zero-mode operator introduced below. We denote by $C_{w\to j}(t)$ the OTOC with source $W=Z_w$ and probe $V=Z_j$, and also consider $V=X_j$ as an alternative probe. To quantify scrambling from the left boundary into the bulk, we define the spatially averaged OTOC
\begin{equation}
\Cbar(t)=\frac{1}{L-6}\sum_{j=3}^{L-4}C_{0\to j}(t),
\label{eq:Cbar}
\end{equation}
where the three sites nearest to each boundary are excluded to minimize contributions from the spatially localized edge modes.

\subsection{Zero-mode weight and edge lifetime} 
The spectral function of $Z_0$ contains a narrow low-frequency peak associated with the almost strong zero mode, superimposed on a broad continuum at frequencies of order $J$~\cite{kemp2017,else2017,yates2020dyn}. We quantify the weight of this low-frequency contribution by $q(\omega_c)$, defined as the fraction of spectral weight within the frequency window $|\omega|<\omega_c$ [defined in  Eq.~\eqref{eq:qwc}]. Using this spectral window, we construct a normalized zero-mode operator,
\begin{equation}
\hat\Psi=\frac{1}{\sqrt q}\sum_{|E_m-E_n|<\omega_c}|m\rangle\langle m|Z_0|n\rangle\langle n|,
\label{eq:Psi}
\end{equation}
where $|m\rangle$ and $E_m$ are the eigenstates and eigenvalues of $H$, respectively. Throughout, we use $\omega_c=0.1J$; this choice is justified in Sec.~\ref{sec:memory}.

A conventional measure of boundary memory is the half-life $\thalf$, defined by $\Aedge(\thalf)=1/2$. However, this definition does not isolate the decay of the slow component: it depends on both the zero-mode weight and the decay of the remaining spectral contribution. In particular, when $q<1/2$, the autocorrelation can cross the half-height threshold during its initial transient, before the slow component has appreciably decayed. We therefore characterize the lifetime of the protected boundary memory by the edge lifetime $\tedge$, defined through
\begin{equation}
\Aedge(\tedge)=\frac{q}{2}.
\label{eq:tau}
\end{equation}
This definition measures the time at which the autocorrelation has fallen to half the zero-mode plateau weight and thus provides a more appropriate measure of the lifetime of the slowly decaying edge contribution.

We compute these quantities numerically for system sizes up to $L=18$. For $L\le14$ we use exact diagonalization (ED), which gives the edge memory and the zero-mode weight exactly at all times. OTOCs are obtained by dynamical typicality~\cite{bartsch2009,elsayed2013,xu2024tutorial}, in which the trace is replaced by an average over random states, and for $L=16$ and 18 time evolution is carried out with a Chebyshev expansion~\cite{talezer1984,weisse2006}. Details of these methods and their accuracy are given in Appendix~\ref{app:methods}.
\section{Results}
\label{sec:results}

\subsection{Bulk chaos}

\begin{figure}
\includegraphics[width=\columnwidth]{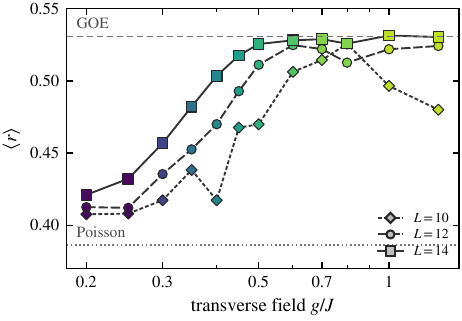}
\caption{Mean adjacent-gap ratio $\langle r\rangle$ versus transverse field $g$ for $L=10$, 12, and 14, with $J=1$ and $K=-0.3$. The ratio is computed in the parity sector $P=+1$ and averaged over the two reflection sectors. Dashed and dotted lines indicate the GOE and Poisson values, respectively.}
\label{fig:levels}
\end{figure} 
We first establish the extent to which the bulk exhibits quantum-chaotic behavior in the parameter regime of interest. To this end, we examine the mean adjacent-gap ratio $\langle r\rangle$, a spectral diagnostic of quantum chaos~\cite{oganesyan2007,atas2013}. Given the ordered energy levels $E_n$, we define the consecutive level spacings as $\delta_n=E_{n+1}-E_n$ and the adjacent-gap ratio as
\begin{equation}
r_n=\frac{\min(\delta_n,\delta_{n+1})}{\max(\delta_n,\delta_{n+1})},\qquad
\langle r\rangle=\frac{1}{N_{\rm lev}-2}\sum_{n=1}^{N_{\rm lev}-2}r_n,
\label{eq:r}
\end{equation}
where $N_{\rm lev}$ is the number of levels in the symmetry sector under consideration. The mean ratio approaches $\langle r\rangle_{\rm GOE}\simeq0.531$ for Gaussian orthogonal ensemble (GOE) statistics, characteristic of quantum-chaotic systems with time-reversal symmetry, and $\langle r\rangle_{\rm P}\simeq0.386$ for Poisson statistics, characteristic of integrable systems.

A proper comparison requires resolving all exact unitary symmetries before computing the level statistics, since combining levels from distinct symmetry sectors can introduce artificial level crossings and suppress level repulsion. Our Hamiltonian commutes with the global spin-flip operator $P=\prod_jX_j$ and the reflection operator $R$, which acts as $RZ_jR^\dagger=Z_{L-1-j}$. We therefore calculate the level statistics in the joint symmetry sectors, fixing $P=+1$ and resolving the two reflection eigenvalues $R=\pm1$. The results from the two reflection sectors are then combined using dimension-weighted averages. Since the Hamiltonian is real in the $Z$ basis, each resolved sector has the relevant time-reversal symmetry and is expected to exhibit GOE statistics in the chaotic regime.

Figure~\ref{fig:levels} reveals a clear crossover in the spectral statistics as the transverse field is varied. For $g\gtrsim0.5$, the mean adjacent-gap ratio approaches the GOE prediction, reaching a value within $0.01$ of it for $L=14$. At weaker fields, the deviation from GOE statistics becomes more pronounced, but decreases systematically with increasing system size. For instance, at $g=0.4$, $\langle r\rangle$ increases from $0.417$ for $L=10$ to $0.503$ for $L=14$, while at $g=0.3$ it rises from $0.417$ to $0.457$. This size dependence suggests that the crossover to GOE-like statistics shifts towards weaker fields as the system grows. The departure from GOE behavior at small $g$ is consistent with the approximate conservation of the Ising interaction energy near the classical Ising limit~\cite{else2017,abanin2017}. In this regime, the transverse field mixes configurations through resonant processes that become increasingly effective as the density of near-degenerate configurations grows with system size. Although our finite-size results do not establish the asymptotic spectral statistics at weak fields, they indicate a systematic enhancement of level repulsion with increasing $L$.

\begin{figure*}
\includegraphics[width=\textwidth]{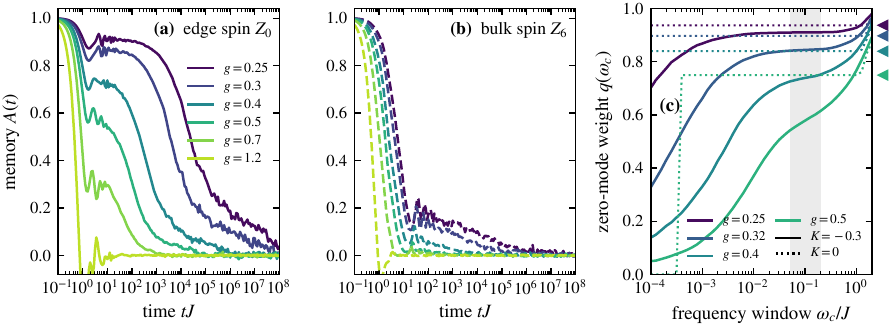}
\caption{Memory and the almost strong zero mode for $L=12$, $J=1$, and $K=-0.3$. (a) Edge-spin memory $\Aedge(t)=\mathrm{Tr}[Z_0(t)Z_0]/D$ and (b) central-spin memory $A_{Z_6}(t)=\mathrm{Tr}[Z_6(t)Z_6]/D$, for several transverse fields $g$. (c) Low-frequency spectral weight $q(\omega_c)$ of $Z_0$, integrated over $|\omega|<\omega_c$, for $K=-0.3$ (solid curves) and for the integrable chain with $K=0$ (dotted curves). Triangles at the right edge mark the exact $K=0$ value $q=1-g^2$, the overlap of $Z_0$ with the strong zero mode~\cite{fendley2012,kemp2017}. The shaded region $0.05\le\omega_c/J\le0.2$ brackets our choice $\omega_c=0.1J$.}
\label{fig:memory}
\end{figure*}

\subsection{Edge memory and the almost strong zero mode}
\label{sec:memory}

We examine whether the slow edge dynamics is accompanied by similarly slow relaxation in the bulk by comparing the autocorrelation of the edge spin with that of the central spin. Figures~\ref{fig:memory}(a) and \ref{fig:memory}(b) reveal a striking separation between the two. For $g\leq 0.5$, the central-spin memory decays rapidly, while the edge memory exhibits a pronounced plateau and relaxes only on much longer timescales. This contrast is particularly evident at $g=0.25$, where the edge memory remains as large as $\Aedge(10)=0.90$ and $\Aedge(10^3)=0.85$. At the same time, the central-spin autocorrelation at $t=10$ has already decreased to $0.29$, $0.09$, and $0.02$ for $g=0.25$, $0.4$, and $0.5$, respectively. The persistence of edge memory despite the substantial relaxation of the bulk demonstrates that the boundary is protected by a mechanism distinct from globally slow dynamics.

This separation disappears at stronger transverse fields. For $g=1.2$, which is comparable to the lowest edge-flip energy $2(J-|K|)=1.4J$, the edge and central-spin autocorrelations decay on comparable timescales. Thus, the long-lived boundary memory at weak transverse fields is associated with an almost strong zero mode, which provides an approximate conservation law localized near the edge~\cite{kemp2017,else2017,yates2020}. We now quantify how much of the edge spin belongs to this slow mode.

To quantify the low-frequency component responsible for this long-lived memory, we examine the spectral weight of the edge operator $Z_0$ within a frequency window $|\omega|<\omega_c$. In the eigenbasis of $H$, with eigenstates $|m\rangle$ and energies $E_m$, each matrix element $\langle m|Z_0|n\rangle$ describes a component of $Z_0$ that oscillates at frequency $\omega_{mn}=E_m-E_n$, since $\langle m|Z_0(t)|n\rangle=e^{i\omega_{mn}t}\langle m|Z_0|n\rangle$. Its contribution to the norm of $Z_0$ is $w_{mn}=|\langle m|Z_0|n\rangle|^2/D$, and these weights sum to one. The weight of the slow component is obtained by summing over all pairs of eigenstates whose energy difference lies within the window,
\begin{equation}
q(\omega_c)=\frac{1}{D}\sum_{|E_m-E_n|<\omega_c}\big|\langle m|Z_0|n\rangle\big|^2 .
\label{eq:qwc}
\end{equation}
The same weights determine the memory, $\Aedge(t)=\sum_{mn}w_{mn}\cos(\omega_{mn}t)$. Once the fast components have dephased ($t\gg1/\omega_c$), and before the slow component itself decays, $\Aedge$ therefore stays close to $q$. 

The integrable chain provides an exact reference. At $K=0$, the strong zero mode commutes with the Hamiltonian, so its entire weight lies at zero frequency, while the remaining part of $Z_0$ oscillates at frequencies above a bulk gap of order $J$. The overlap of the edge spin with the zero mode is known exactly, $q=1-g^2$~\cite{fendley2012,kemp2017}, and $q(\omega_c)$ should therefore be independent of $\omega_c$ throughout this gap.  Our numerical results [dotted lines in Fig.~\ref{fig:memory}(c)] confirm this behavior: $q(\omega_c)$ is flat and equal to $1-g^2$ to better than $10^{-6}$ for all fields shown, and rises only when $\omega_c$ approaches the bulk gap $2(J-g)$, near the right edge of the panel. The procedure thus measures exactly the part of $Z_0$ that belongs to the zero mode.

Breaking integrability changes this picture in two ways. First, the almost strong zero mode acquires a finite lifetime $\tau_{\rm edge}$, so its weight is spread over a peak of width of order $1/\tau_{\rm edge}$; windows narrower than this peak collect only part of it, and $q(\omega_c)$ decreases at small $\omega_c$. Second, the interacting bulk has spectral weight at all frequencies, so the gap fills in and $q(\omega_c)$ grows slowly at large $\omega_c$. A plateau therefore survives only between these two scales, $1/\tau_{\rm edge}\ll\omega_c\ll J$. Figure~\ref{fig:memory}(c) shows that for $K=-0.3$ such a plateau is present at small fields. At $g=0.25$, where the lifetime is longest, it extends over about two decades in $\omega_c$. As $g$ increases, the lifetime decreases, the peak broadens, and the plateau becomes narrower. We fix $\omega_c=0.1J$, inside the shaded interval, and use this value throughout the rest of the paper. Varying $\omega_c$ across this interval changes $q$ by less than $0.015$ for $g\leq0.4$, and by about $\pm0.04$ at $g=0.5$, where the peak and the continuum nearly overlap.

With this choice we find $q=0.91$, $0.84$, $0.74$, and $0.58$ at $g=0.25$, $0.32$, $0.4$, and $0.5$. These values lie below the integrable results $1-g^2$, and the reduction grows with $g$, from $0.03$ at $g=0.25$ to $0.17$ at $g=0.5$, reflecting the loss of exact zero-mode protection. Nevertheless, for $g\leq0.5$ most of the edge spin is still carried by a slow component, which is the origin of the long memory in Fig.~\ref{fig:memory}(a). The decrease of $q$ continues at larger fields. We find $q=0.32$ at $g=0.7$ and $q\approx0.005$ at $g=1.2$, consistent with the absence of a long edge memory at the largest field.

\begin{figure*}
\includegraphics[width=\textwidth]{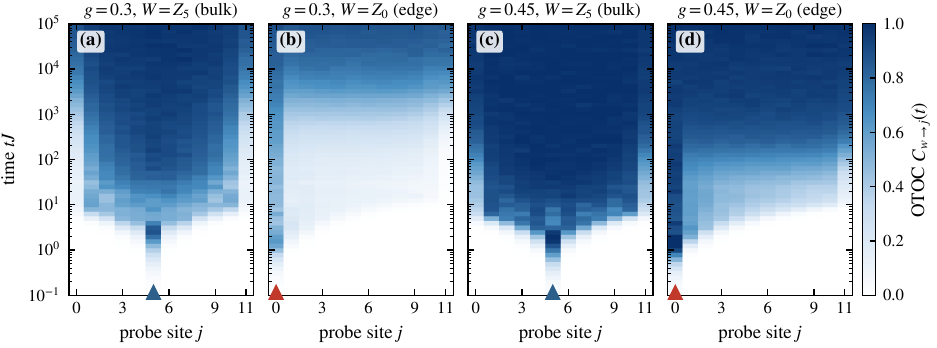}
\caption{Space--time profiles of the OTOC $C_{w\to j}(t)$ for $L=12$, $J=1$, and $K=-0.3$ at infinite temperature, shown on a logarithmic time axis. Triangles indicate the source site. (a), (c) Bulk source $w=5$; (b), (d) edge source $w=0$; $g=0.3$ in (a), (b) and $g=0.45$ in (c), (d). A bulk source produces conventional light-cone spreading, whereas an edge source generates a slow, nearly spatially uniform growth across the bulk, preceded by a weak propagating front associated with the fast component of $Z_0$.}
\label{fig:maps}
\end{figure*}
\subsection{Space--time structure of edge scrambling}
\label{sec:maps}

The space--time structure of the OTOC reveals a striking difference between information spreading from the bulk and from the boundary. Figures~\ref{fig:maps}(a) and \ref{fig:maps}(c) show the dynamics for a bulk source at $w=5$. Since $L=12$ is even, the chain has no single central site; the source at $w=5$ lies five sites from the left boundary and six sites from the right boundary. This one-site asymmetry does not affect the conclusions below. In this case, the OTOC develops a clear light cone: scrambling reaches progressively more distant sites as time increases, and most of the chain is affected by $t\approx10$. Averaged over the bulk sites $3\le j\le8$, the OTOC already reaches $0.86$ at $t=100$ for $g=0.3$ [Fig.~\ref{fig:maps}(a)] and $0.99$ for $g=0.45$ [Fig.~\ref{fig:maps}(c)], corresponding to the dark region of the color scale. Thus, although bulk scrambling is not instantaneous, particularly at the weaker field, it proceeds efficiently on timescales much shorter than the edge-memory lifetime. The two boundary sites remain strongly suppressed, indicating that information initially placed in the bulk has difficulty reaching the edges. For the real Hamiltonian in Eq.~\eqref{eq:H}, the infinite-temperature OTOC is symmetric under interchange of source and probe. The suppression of scrambling at the boundary therefore reflects the same dynamical restriction that inhibits information from escaping an edge source.

The dynamics changes qualitatively when the source is placed at the boundary, as shown in Figs.~\ref{fig:maps}(b) and \ref{fig:maps}(d). Rather than developing a conventional light cone, the edge-to-bulk OTOC grows slowly over a broad spatial region. For $g=0.3$, the bulk-averaged OTOC is only $\Cbar=0.17$ at $t=100$, but increases to $0.76$ by $t=10^4$. Despite this slow temporal evolution, the spatial profile becomes remarkably uniform: for $t\ge10^3$, the OTOCs at $j=3$ and $j=8$ differ by less than $0.02$. In other words, once the edge information begins to enter the bulk, distant sites respond on nearly the same timescale, rather than being reached sequentially through a propagating front.

These results distinguish two stages of edge-to-bulk scrambling. A small fraction of the initial edge operator can spread ballistically at early times, but the subsequent growth across the bulk is governed primarily by the slow release of the long-lived boundary component. The bulk therefore does not set the dominant timescale through its butterfly velocity; instead, the boundary acts as a bottleneck that controls when information becomes available for scrambling in the bulk. We refer to this regime as \emph{bottlenecked scrambling}.

\begin{figure*}
\includegraphics[width=\textwidth]{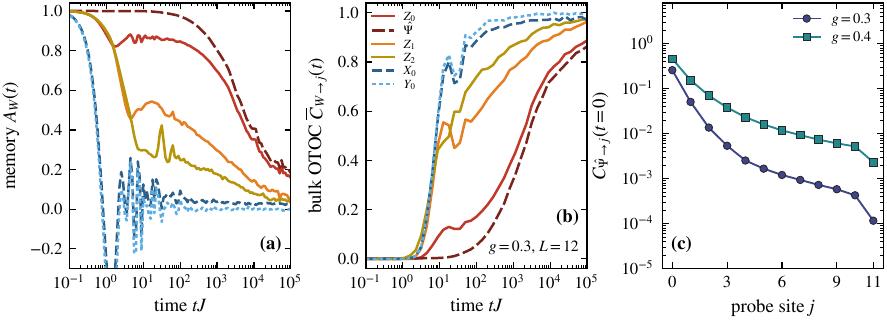}
\caption{Operator dependence of edge memory and scrambling for $L=12$, $J=1$, and $K=-0.3$. (a) Autocorrelation $A_W(t)$ and (b) bulk-averaged OTOC $\overline{C}_{W\to j}(t)$, averaged over the probes $Z_j$ with $3\le j\le8$, at $g=0.3$ for the source operators $W=Z_0$, $X_0$, $Y_0$, $Z_1$, $Z_2$, and the zero-mode operator $\hat\Psi$ of Eq.~\eqref{eq:Psi}. (c) Equal-time OTOC $C_{\hat\Psi\to j}(0)$ with the probes $Z_j$ as a function of the probe site $j$, for $g=0.3$ and $0.4$.}
\label{fig:ops}
\end{figure*}
\subsection{Operator selectivity of edge protection}
\label{sec:ops}

In the integrable chain, it is known which boundary operators are protected. At $K=0$, the edge spin $Z_0$ overlaps with the strong zero mode, whereas the transverse operators $X_0$ and $Y_0$ do not~\cite{kitaev2001,fendley2012,alicea2012}. It is not obvious that this selectivity survives in the nonintegrable chain, where the zero mode is only approximately conserved. If the long edge memory indeed originates from an almost conserved edge mode, however, only operators that overlap with this mode should be protected, not every operator at the boundary~\cite{kemp2017,else2017,fendley2016}. We therefore expect $Z_0$ to keep its memory, while $X_0$ and $Y_0$ should lose it quickly, even though all three operators act on the same site.

To isolate the protected part of the edge spin directly, we also consider the zero-mode operator $\hat\Psi$ of Eq.~\eqref{eq:Psi}, obtained from $Z_0$ by keeping only its matrix elements between eigenstates with nearly equal energies. These are exactly the components of $Z_0$ that oscillate slowly, so $\hat\Psi$ is the slow part of the edge spin, and the factor $1/\sqrt{q}$ normalizes it. In the integrable limit, this filtering returns the exact strong zero mode $\Psi_0$ of Eq.~\eqref{eq:Psi0}; for $K\neq0$ it provides an explicit construction of the almost strong zero mode. We compare $\hat\Psi$ with the edge operators $Z_0$, $X_0$, and $Y_0$ and with the near-edge operators $Z_1$ and $Z_2$, all at $g=0.3$.

The memories confirm the expectation from the integrable chain [Fig.~\ref{fig:ops}(a)]. The zero-mode operator $\hat\Psi$ is almost perfectly conserved, with a memory that remains above $0.97$ up to $t=100$. The edge spin $Z_0$ retains a large but smaller memory, $A_{Z_0}\simeq0.85$ at $t=100$, close to its zero-mode weight $q=0.87$. In contrast, $X_0$ and $Y_0$ lose their memory almost immediately, falling below $0.13$ by $t=10$, as expected for operators without overlap with the zero mode. The neighboring longitudinal operators $Z_1$ and $Z_2$ show intermediate behavior, with partial plateaus at $A_{Z_1}=0.44$ and $A_{Z_2}=0.25$ at $t=100$. Slow dynamics thus extends slightly beyond the edge site, without these operators being zero modes themselves.

The same hierarchy appears, more sharply, in the spreading of information into the bulk [Fig.~\ref{fig:ops}(b)]. Information carried by $\hat\Psi$ remains almost entirely at the boundary: its bulk-averaged OTOC reaches only $0.004$ at $t=10$ and $0.06$ at $t=100$. For $Z_0$, the OTOC reaches $0.10$ at $t=10$ and $0.17$ at $t=100$. The difference from $\hat\Psi$ comes from the finite-frequency component of $Z_0$, with weight $1-q\simeq0.13$, which leaks into the bulk at early times and which $\hat\Psi$ removes by construction. The transverse operators, by contrast, scramble like bulk operators: the OTOCs of $X_0$ and $Y_0$ reach approximately $0.63$ and $0.65$ at $t=10$ and exceed $0.88$ by $t=100$. Although $Z_1$ and $Z_2$ retain partial memory, their OTOCs already reach approximately $0.5$ within $t\approx10$, so a residual memory does not by itself imply an equally strong suppression of scrambling. Protection against scrambling is therefore controlled by the overlap with the almost strong zero mode, not by the location of the operator alone, and the selectivity known from the integrable chain survives the breaking of integrability.

The zero-mode operator is concentrated at the boundary. To see where $\hat\Psi$ lives, we compute its OTOC with the local probes $Z_j$ at $t=0$ [Fig.~\ref{fig:ops}(c)]. Writing an operator as a sum of Pauli strings, $\hat\Psi=\sum_S c_S S$, each string either commutes with $Z_j$, in which case it does not contribute to the commutator, or anticommutes with it, in which case $[S,Z_j]=2SZ_j$. The equal-time OTOC is therefore twice the weight of the operator on the anticommuting strings,
\begin{equation}
C_{\hat\Psi\to j}(0)=\frac{1}{2D}\,\mathrm{Tr}\big([\hat\Psi,Z_j]^\dagger[\hat\Psi,Z_j]\big)=2\sum_{S:\,\{S,Z_j\}=0}|c_S|^2 .
\label{eq:OTOCweight}
\end{equation}
This quantity is large if $\hat\Psi$ has weight near site $j$ and small otherwise. For $g=0.3$, it drops by a factor of about $48$ from $j=0$ to $j=3$.

For comparison, we evaluate Eq.~\eqref{eq:OTOCweight} for the exact zero mode of Eq.~\eqref{eq:Psi0}~\cite{fendley2012,kemp2017}. Its strings are $X_0\cdots X_{k-1}Z_k$, with weights $(1-g^2)g^{2k}$. Such a string anticommutes with $Z_j$ only if it contains $X_j$, that is, only if $k>j$. Summing the geometric series gives
\begin{equation}
C_{\Psi_0\to j}(0)=2(1-g^2)\sum_{k=j+1}^{\infty}g^{2k}=2g^{2(j+1)},
\label{eq:Psiprofile}
\end{equation}
up to corrections of order $g^{2L}$ in a finite chain. This profile drops by a factor $C_{\Psi_0\to0}(0)/C_{\Psi_0\to3}(0)=2g^{2}/2g^{8}=g^{-6}\simeq1.4\times10^3$ from $j=0$ to $j=3$. The decay at $K=-0.3$ is slower for two reasons: the integrability-breaking interaction spreads the edge mode over a few more sites, and the filtering procedure adds a small tail. At $g=0.4$ the drop is only a factor of $12$, so the mode is less localized at the larger field.

Part of this tail is an artifact of the filtering. Keeping only the frequencies $|\omega|<\omega_c$ in Eq.~\eqref{eq:Psi} is equivalent to averaging the evolving edge spin over time,
\begin{equation}
\sqrt{q}\,\hat\Psi=\int_{-\infty}^{\infty}dt\,\frac{\sin(\omega_c t)}{\pi t}\,Z_0(t),
\label{eq:Psitime}
\end{equation}
where the weight function extends over times of order $1/\omega_c$. In simple terms, $\hat\Psi$ is the part of $Z_0$ that survives when $Z_0(t)$ is averaged over this time window. During this window, however, a small part of the edge spin has already leaked into the bulk, and this leaked part is included in the average. The smaller $\omega_c$, the longer the averaging time, and the larger this artificial tail. If, on the other hand, $\omega_c$ is too large, the filter also starts to include fast bulk components of $Z_0$, which again broadens the profile. The tail is therefore smallest for $\omega_c$ inside the plateau of $q(\omega_c)$, where our choice $\omega_c=0.1J$ lies. The strong concentration of $\hat\Psi$ on the first few sites also justifies excluding the first three sites from the bulk average in Eq.~\eqref{eq:Cbar}.

\subsection{Time-doubling law}
\label{sec:leak}

\begin{figure*}[t]
\includegraphics[width=\textwidth]{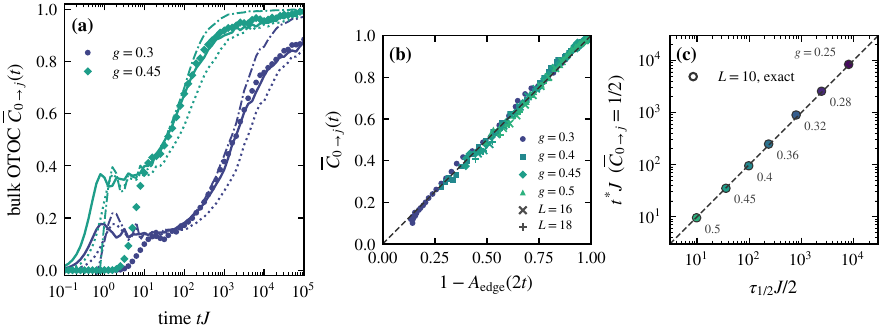}
\caption{Time-doubling law. (a) Bulk-averaged OTOC $\Cbar(t)$ (symbols, $L=12$) compared with $1-\Aedge(2t)$ (solid lines), $1-\Aedge(t)^2/q$ (dash-dotted lines), and $1-\Aedge(t)$ (dotted lines) for $g=0.3$ and 0.45. (b) Parametric plot of $\Cbar(t)$ against $1-\Aedge(2t)$ for $t\ge10$ ($L=12$, filled symbols), and of the central-probe OTOC $C_{0\to L/2}(t)$ for $t\ge20$ ($L=16$, crosses; $L=18$, plus signs). Colors encode $g$, and the dashed diagonal is the parameter-free prediction. (c) Time $t^*$ at which the OTOC, averaged over the central probes $4\le j\le6$, reaches $1/2$, plotted against half the memory half-life, for $0.25\le g\le0.5$ (exact ED, $L=10$); each point is labelled by its value of $g$, and the dashed line marks $t^*=\thalf/2$.}
\label{fig:leak}
\end{figure*}
The long-lived memory of the edge spin and the slow growth of the bulk OTOC suggest that information spreads from the boundary through a gradual leakage of the almost conserved edge mode. This picture also implies that the growth of the OTOC should be related to the decay of the edge autocorrelation. We now derive a quantitative relation between these two quantities.

We decompose the time-evolved edge spin into two contributions. After a short transient, during which the fast, finite-frequency component of $Z_0$ spreads away from the boundary, we write
\begin{equation}
Z_0(t)=\mathcal{E}(t)+\mathcal{B}(t),
\label{eq:EB}
\end{equation}
where $\mathcal{E}(t)$ denotes the component that remains localized near the edge, while $\mathcal{B}(t)$ describes the component that has spread into the bulk. Since unitary evolution preserves the operator norm, the weights of these two components satisfy
\[
\frac{\mathrm{Tr}[\mathcal{E}(t)^2]}{D}
+
\frac{\mathrm{Tr}[\mathcal{B}(t)^2]}{D}
=1.
\]
A probe sufficiently far from the boundary commutes with $\mathcal{E}(t)$, since this component remains localized near the edge. Consequently, the commutator in Eq.~\eqref{eq:C}, and hence the bulk OTOC, receives contributions only from $\mathcal{B}(t)$. The OTOC is therefore determined by the weight that has spread into the bulk, or, equivalently, by the weight that remains localized at the edge.

The remaining edge weight is not directly given by the autocorrelation $\Aedge(t)$, which measures the overlap of $Z_0(t)$ with its initial form. However, time-reversal symmetry allows us to relate this weight to the autocorrelation at twice the time. To show this, we expand the evolved operator in the Pauli-string basis,
\[
Z_0(t)=\sum_S c_S(t)S,
\]
where the coefficients $c_S(t)$ are real because $Z_0(t)$ is Hermitian, and satisfy $\sum_S c_S(t)^2=1$. Since both the Hamiltonian and $Z_0$ are real in the $Z$ basis, complex conjugation reverses the direction of time:
\[
Z_0(t)^*=Z_0(-t).
\]
The Pauli matrices $X$ and $Z$ are real, whereas $Y$ is purely imaginary. Thus, a Pauli string transforms under complex conjugation as
\[
S^*=(-1)^{n_Y(S)}S,
\]
where $n_Y(S)$ denotes the number of $Y$ operators in the string. Using this property and the cyclicity of the trace, we can express the edge memory at time $2t$ as
\begin{equation}
\Aedge(2t)=\frac{1}{D}\,\mathrm{Tr}\big[Z_0(-t)Z_0(t)\big]
=\sum_S(-1)^{n_Y(S)}\,c_S(t)^2.
\label{eq:TR}
\end{equation}
This identity is exact. It shows that the memory at time $2t$ is obtained by summing the weights of all Pauli strings in $Z_0(t)$, with a positive sign for strings containing an even number of $Y$ operators and a negative sign for those containing an odd number.

We now consider how the two components in Eq.~\eqref{eq:EB} contribute to this sum. On the slow timescale, the edge component $\mathcal{E}(t)$ is dominated by the zero-mode operator $\hat\Psi$ defined in Eq.~\eqref{eq:Psi}. Since the eigenvectors of the real Hamiltonian can be chosen real, $\hat\Psi$ is also real. Its Pauli-string expansion therefore contains only strings with an even number of $Y$ operators. The edge component consequently contributes its full weight, $\mathrm{Tr}[\mathcal{E}(t)^2]/D$, to Eq.~\eqref{eq:TR}, with a positive sign. The bulk component behaves differently. Once it has been scrambled by the chaotic bulk, its weight is distributed over a large number of Pauli strings, with even and odd numbers of $Y$ operators occurring with approximately equal probability. Their contributions to Eq.~\eqref{eq:TR} then cancel. Under these conditions, the memory at time $2t$ is approximately equal to the weight that remains near the edge at time $t$:
\begin{equation}
\frac{1}{D}\,\mathrm{Tr}\big[\mathcal{E}(t)^2\big]
\simeq \Aedge(2t).
\label{eq:Enorm}
\end{equation}
It follows from norm conservation that the weight leaked into the bulk is
\[
\frac{\mathrm{Tr}[\mathcal{B}(t)^2]}{D}
\simeq 1-\Aedge(2t).
\]

We next connect this leaked weight to the OTOC. Consider a single-site Pauli probe $V_j$. Each Pauli string in $\mathcal{B}(t)$ either commutes or anticommutes with $V_j$. Only the anticommuting strings contribute to the commutator, and the OTOC defined in Eq.~\eqref{eq:C} is twice the total weight of these strings. In a fully scrambled operator, the four possible single-site Pauli operators, $I$, $X$, $Y$, and $Z$, occur with equal weight. Since two of them anticommute with any given single-site Pauli probe, half of the leaked weight contributes to the OTOC. The OTOC therefore equals the total weight that has leaked into the bulk, giving
\begin{equation}
\Cbar(t)\simeq1-\Aedge(2t).
\label{eq:doubling}
\end{equation}
Since the same argument applies to each bulk probe, the relation also holds for the bulk-averaged OTOC. This is the time-doubling law. The detailed derivation is provided in Appendix~\ref{app:derivation}.

The factor of two in Eq.~\eqref{eq:doubling} follows from the distinction between the edge memory and the leaked weight. The autocorrelation $\Aedge(t)$ measures the overlap of the evolved edge operator with its initial form, whereas the OTOC measures the weight that has spread into the bulk. The latter is determined by the square of the amplitude that remains at the edge. Squaring an exponentially decaying amplitude doubles its decay rate. More generally, suppose that the slow edge memory is a sum of exponential contributions,
\[
\Aedge(t)\simeq\sum_k q_k e^{-\Gamma_k t}.
\]
The corresponding edge amplitudes at time $t$ are proportional to $\sqrt{q_k}e^{-\Gamma_k t}$, so their squared contributions give $\sum_k q_k e^{-2\Gamma_k t}=\Aedge(2t)$. Importantly, the time-doubling relation does not require the memory to decay as a single exponential, or even as a sum of exponentials. Its validity follows from the time-reversal identity in Eq.~\eqref{eq:TR}, together with the separation between the localized edge component and the scrambled bulk component.

This relation has a natural counterpart in systems with conserved quantities. In such systems, the part of an operator associated with a conserved density cannot be scrambled, and the OTOC develops slow tails that reflect the decay of its overlap with the conserved quantities~\cite{khemani2018,rakovszky2018,balachandran2021,balachandran2023}. Here, the almost conserved edge mode plays a similar role. The key difference is that the protected component is localized at the boundary and gradually loses its weight by spreading into the bulk. The OTOC therefore grows as this edge-localized component decays.

The time-doubling law contains no adjustable parameters, but its accuracy depends on several physical conditions. It relies on time-reversal symmetry, which gives the exact identity in Eq.~\eqref{eq:TR}. It also requires a separation of timescales: the edge must release its weight more slowly than the bulk scrambles it, so that the leaked component is already scrambled when probed. In other words, the edge-leakage rate must be much smaller than the bulk scrambling rate. Near the Ising limit, the bulk scrambling rate can itself be much smaller than $J$. Finally, the probe must be located beyond the localization length of the zero mode, so that it does not directly overlap with the edge component. The relation is expected to hold only after the leaked information has had sufficient time to reach the probe.

We now compare these predictions with the numerical results in Fig.~\ref{fig:leak}, starting with the time dependence. Figure~\ref{fig:leak}(a) shows that, after the leaked signal reaches the bulk, the bulk-averaged OTOC closely follows $1-\Aedge(2t)$ for both fields shown, $g=0.3$ and $g=0.45$. To establish the importance of the time-doubling factor, we also compare the data with two alternative expressions. The quantity $1-\Aedge(t)$, which identifies the OTOC with the memory lost by time $t$, systematically underestimates the OTOC. In contrast, $1-\Aedge(t)^2/q$, which follows from the time-doubling law for a single-exponential decay, overestimates it, particularly at late times and smaller $g$. These expressions are not arbitrary; they provide bounds implied by the time-doubling law. When the slow memory is a sum of exponential contributions, monotonicity of the decay and the Cauchy--Schwarz inequality give (Appendix~\ref{app:derivation})
\begin{equation}
1-\Aedge(t)\;\leq\;\Cbar(t)\;\leq\;1-\Aedge(t)^2/q\;\leq\;1-\Aedge(t)^2.
\label{eq:band}
\end{equation}
The lower bound shows that the OTOC is at least as large as the memory lost by time $t$. The upper bound is saturated only when all contributions decay at the same rate, corresponding to a single exponential. The numerical data lie clearly below this upper bound, showing that the slow edge memory does not decay as a single exponential, consistent with the lifetime analysis in Sec.~\ref{sec:tau}. Among the three expressions, only the time-doubling law accurately captures the numerical results without assuming a particular form for the decay.

We next test the validity of the relation across different fields and system sizes. Figure~\ref{fig:leak}(b) presents a parametric plot of the OTOC against $1-\Aedge(2t)$. This provides a direct test of the relation, since deviations appear as departures from the diagonal. Without any fitting parameters, the data for $L=12$ and for the central probes at $L=16$ and $18$ collapse onto the diagonal. The remaining deviations are at the level of a few percent, comparable to or slightly larger than the statistical uncertainty of the numerics. They are consistent with the limitations of the assumptions underlying the derivation. In particular, the deviations are largest at the weakest field, where bulk scrambling is slower and the separation of timescales is less pronounced. In this regime, recently leaked weight may not yet be fully scrambled. Deviations are also more noticeable for probes close to the opposite boundary, which approach the predicted relation only at later times. The relation also holds for $X_j$ probes, with somewhat lower accuracy, and with comparable accuracy when the zero-mode operator $\hat\Psi$ is used as the source, as expected from the derivation, which requires only that the probe is a single-site Pauli operator.

Finally, the time-doubling law gives a direct relation between the scrambling time and the edge-memory lifetime, which we test in Fig.~\ref{fig:leak}(c). We define $t^*$ as the time at which the bulk OTOC reaches one half, and $\thalf$ as the time at which the edge memory reaches one half. According to Eq.~\eqref{eq:doubling}, the condition $\Cbar(t^*)=1/2$ implies $\Aedge(2t^*)=1/2$. Therefore,
\begin{equation}
t^*\simeq\thalf/2.
\label{eq:half}
\end{equation}
This relation does not depend on the functional form of the decay, provided that the memory reaches one half on the slow timescale, which requires $q>1/2$. Figure~\ref{fig:leak}(c) confirms this prediction for scrambling times ranging from about $10$ to nearly $10^4\,J^{-1}$. The data are obtained from exact diagonalization for $L=10$, using the full trace without random-state sampling. Throughout the range $0.25\leq g\leq0.5$, the ratio $t^*/\thalf$ remains close to $1/2$. Since these results contain no statistical sampling noise, the small deviations reflect the approximate nature of the time-doubling law itself. At larger fields, where $q<1/2$, the edge memory reaches one half during the initial fast decay, before the slow zero-mode component has decayed appreciably. In this regime, $\thalf$ no longer characterizes the lifetime of the edge mode, and the relation in Eq.~\eqref{eq:half} no longer applies.
\subsection{Lifetime of the slow component}
\label{sec:tau}

\begin{figure*}[t]
\includegraphics[width=\textwidth]{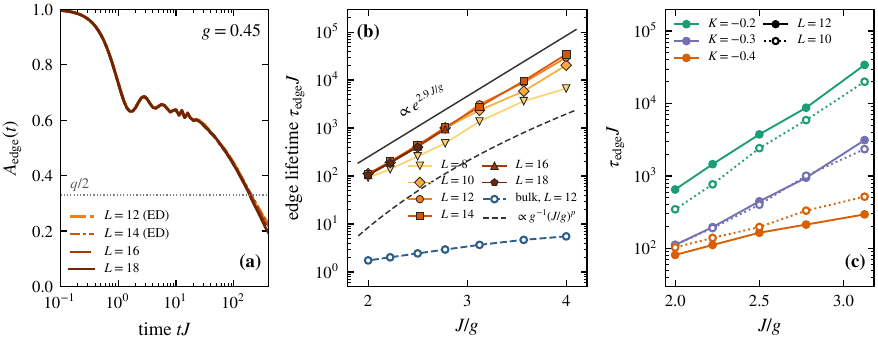}
\caption{Lifetime of the slow component for $J=1$. (a) Edge memory $\Aedge(t)$ at $g=0.45$ and $K=-0.3$ for $L=12$ and 14 (ED) and $L=16$ and 18 (typicality). The dotted line marks $q/2$, with $q=0.66$ obtained at $\omega_c=0.1J$. (b) Edge lifetime $\tedge$ at $K=-0.3$ as a function of $J/g$ for $L=8$--18 (darker shading for larger $L$), together with the half-life of the central spin for $L=12$ (open circles). The solid line is an exponential fit, $\tedge\propto e^{cJ/g}$, and the dashed line a fit to the form $\tedge\propto g^{-1}(J/g)^p$ of Ref.~\cite{kemp2020}; both fits use the data for $g\ge0.28$ and are shifted vertically for clarity. (c) Edge lifetime $\tedge$ as a function of $J/g$ for $K=-0.2$, $-0.3$, and $-0.4$, for $L=12$ (filled symbols, solid lines) and $L=10$ (open symbols, dotted lines).}
\label{fig:tau}
\end{figure*}

\begin{table}[t]
\caption{Zero-mode weight $q$ and edge lifetime $\tedge J$ [Eq.~\eqref{eq:tau}] at infinite temperature. Values for $L\le14$ are from ED, those for $L=16$ and 18 from dynamical typicality; a dash marks parameters beyond the accessible time window. For $g\le0.28$, the lifetimes still increase with $L$; those for $g=0.25$ are lower bounds for the thermodynamic limit.}
\label{tab:tau}
\begin{ruledtabular}
\begin{tabular}{ccrrrrrr}
$g/J$ & $q$ & $L=8$ & 10 & 12 & 14 & 16 & 18 \\
\colrule
0.25 & 0.911 & 6672 & 20389 & 29662 & 34559 & -- & -- \\
0.28 & 0.885 & 3650 & 5886 & 8860 & 9633 & -- & -- \\
0.32 & 0.845 & 1376 & 2350 & 3131 & 2786 & -- & -- \\
0.36 & 0.796 & 482 & 1012 & 948 & 1038 & 948 & -- \\
0.40 & 0.739 & 257 & 402 & 448 & 436 & 412 & 407 \\
0.45 & 0.660 & 139 & 192 & 197 & 200 & 193 & 188 \\
0.50 & 0.579 & 92 & 112 & 112 & 110 & 106 & 105 \\
\end{tabular}
\end{ruledtabular}
\end{table}

The long-lived edge memory is expected to persist for increasingly long times as the transverse field \(g\) and the next-nearest-neighbor coupling \(K\) become weaker. Previous works have investigated this behavior and derived the asymptotic scaling of the edge lifetime in the regime of small \(g\) and \(|K|\)~\cite{kemp2017,else2017,kemp2020}. Here, we examine its quantitative behavior in the parameter regime accessible to numerics, focusing on its dependence on \(g\) and \(K\), its convergence with system size, and its consistency with the prediction of Kemp, Yao, and Laumann for this model~\cite{kemp2020}.

We begin by examining the system-size dependence of the edge lifetime. In a finite chain, the edge mode can be influenced by the opposite boundary, making it important to establish convergence with increasing \(L\). Figure~\ref{fig:tau}(a) shows the edge memory at \(g=0.45\) for system sizes ranging from \(L=12\) to \(18\). The curves overlap within the statistical uncertainty, indicating that the lifetime is already converged at this field. The lifetimes obtained for the different fields and system sizes are listed in Table~\ref{tab:tau}. For $g\geq0.28$, the results for the two largest available system sizes agree to within about one tenth of their value, suggesting that finite-size effects are relatively small. At $g=0.28$, however, the comparison is limited to $L\leq14$, and the lifetime still increases slowly with system size; its value may therefore be somewhat underestimated. The finite-size effects are more pronounced at $g=0.25$, where the lifetime increases from $20389$ at $L=10$ to $34559$ at $L=14$. These values should thus be regarded as lower bounds. Despite these limitations, the edge lifetime is much longer than the characteristic bulk timescales. As shown in Fig.~\ref{fig:tau}(b), it exceeds the half-life of a spin at the center of the chain by up to approximately four orders of magnitude.

We next investigate the dependence of the lifetime on the transverse field. As shown in Fig.~\ref{fig:tau}(b), $\tedge$ increases rapidly as $g$ decreases, growing by more than two orders of magnitude between $g=0.5$ and $g=0.25$. For this model, Kemp, Yao, and Laumann predicted a power-law dependence on $J/g$~\cite{kemp2020},
\begin{equation}
\tedge\sim\frac{1}{g}\Big(\frac{J}{g}\Big)^{p},
\qquad
p=c_{\rm KYL}\,\frac{J}{|K|},
\label{eq:KYL}
\end{equation}
where $c_{\rm KYL}$ is a constant of order one that was not specified in Ref.~\cite{kemp2020}. Fitting this form to our data for $g\geq0.28$ gives $p=6.9\pm0.4$, corresponding to $c_{\rm KYL}\approx2.1$ [dashed line in Fig.~\ref{fig:tau}(b)]. However, the fit does not fully capture the field dependence: on a logarithmic scale, the data show a clear curvature relative to the fitted power law. This behavior is also reflected in the local exponent, obtained by fitting over narrower field intervals. It increases from $5.7$ for $0.36\leq g\leq0.5$ to $8.2$ for $0.28\leq g\leq0.36$. Thus, a single power law does not adequately describe the entire field range, and the uncertainty obtained from the global fit does not account for this systematic variation. By comparison, a simple exponential form, $\tedge\propto e^{cJ/g}$, with $c=2.9\pm0.1$, describes the data considerably better [solid line in Fig.~\ref{fig:tau}(b)]. The typical deviation of $\ln\tedge$ from this fit is $0.04$, compared with $0.15$ for the power-law fit.

This behavior is not in contradiction with the prediction of Ref.~\cite{kemp2020}, which concerns the asymptotic regime where both $g$ and $|K|$ are much smaller than $J$. Our data are obtained at $|K|/J=0.3$ and cover $2\lesssim J/g\lesssim3.6$, placing them in a crossover regime rather than the asymptotic limit. In this regime, decreasing $g$ increases the number of spin flips required to transfer energy from the edge to the bulk, leading to a rapid growth of the lifetime that is not yet captured by a fixed power law. Over the accessible field range, an exponential dependence, the nearly exponential prethermal bound of Ref.~\cite{else2017}, and the asymptotic power law cannot be clearly distinguished. We therefore interpret $c$ as an effective exponent describing the field range studied here, rather than as an asymptotic scaling exponent. Reaching the asymptotic regime would require substantially longer lifetimes than can be accessed by exact diagonalization, motivating the use of Krylov-space methods such as those developed in Refs.~\cite{yates2020,yates2020dyn}.

The time dependence of the edge memory provides further insight into the underlying decay processes. If the edge mode relaxed through a single decay channel, its memory would follow a single exponential, giving a constant logarithmic derivative. Instead, we find that this derivative decreases with time, indicating a progressively slower decay and the development of a long tail in $\Aedge$. This nonexponential behavior also explains why the single-exponential expression $1-\Aedge(t)^2/q$ fails to describe the OTOC in Sec.~\ref{sec:leak}. A possible explanation is that the edge spin can lose its energy through two distinct channels with different energy costs and, consequently, different decay rates. We now examine the origin of these channels.

The two channels arise from the different energy costs associated with flipping the edge spin. For a fixed configuration of its neighboring spins, the energy change is
\begin{equation}
\Delta E_0=2z_0(Jz_1-Kz_2),
\label{eq:dE0}
\end{equation}
where $z_j=\pm1$ denotes the spin configuration in the $Z$ basis. Consider an edge spin aligned with its nearest neighbor, so that $z_0=z_1$. For $K<0$, flipping this spin costs $2(J+|K|)=2.6J$ when the second neighbor points in the same direction, and $2(J-|K|)=1.4J$ when it points in the opposite direction. The latter is therefore the lower-energy channel and is expected to dominate the edge relaxation.

For the edge spin to flip, the corresponding energy must be absorbed by the rest of the chain. This is strongly constrained by the energy structure of the bulk. A spin flip away from either boundary changes the nearest-neighbor Ising energy by $0$ or $\pm4J$, whereas the edge-flip energy $\Delta E_0$ is an odd multiple of $2J$. Thus, bulk spin flips cannot compensate the energy required for an isolated edge flip through the nearest-neighbor interaction alone. A simultaneous flip at the opposite edge could, in principle, compensate this mismatch, since that flip also costs an odd multiple of $2J$. However, such a process requires the two boundaries to communicate and is exponentially suppressed with increasing $L$, making it a finite-size effect. At $K=0$, the mismatch cannot be compensated by bulk processes. This is the origin of the exact strong zero mode in the integrable chain, for which the edge spin retains its memory indefinitely. When $K\neq0$, the next-nearest-neighbor interaction introduces energy changes in multiples of $2|K|$, allowing the mismatch to be compensated. However, this requires a high-order process involving the rearrangement of approximately $J/|K|$ domain walls~\cite{kemp2020}. Since each rearrangement requires an action of the transverse field, the amplitude acquires a factor of order $g/J$. This provides the physical origin of the large exponent $p\propto J/|K|$ in Eq.~\eqref{eq:KYL}. It also leads to two expectations: the lower-energy channel, with cost $2(J-|K|)$, should dominate the leakage, and increasing $|K|$ should reduce the number of rearrangements required to compensate the energy mismatch, thereby decreasing the exponent.

We test the latter expectation in Fig.~\ref{fig:tau}(c), where we compare the lifetimes for $K=-0.2$, $-0.3$, and $-0.4$. For $L=12$ and $2\leq J/g\lesssim3.1$, the slopes of $\ln\tedge$ as a function of $J/g$ are $3.47$, $2.94$, and $1.14$, respectively. Thus, the lifetime grows more slowly as $g$ decreases when $|K|$ is larger, consistent with the expected trend. A quantitative test of the predicted scaling $p\propto J/|K|$, however, is more difficult. Converting the measured slopes to the power-law form in Eq.~\eqref{eq:KYL} gives $p|K|/J=1.5$ for $K=-0.2$ and $1.9$ for $K=-0.3$. These values would be equal if $p$ were exactly proportional to $J/|K|$. Instead, the corresponding exponents differ by a factor of approximately $1.2$, compared with the predicted factor of $1.5$. The value $1.9$ for $K=-0.3$ also differs from the estimate $c_{\rm KYL}\approx2.1$ obtained above using all system sizes and a wider range of fields, illustrating the systematic uncertainty in extracting the exponent.

The estimates for the other two couplings are subject to additional limitations. At $K=-0.2$, the system lies closer to the integrable point $K=0$, where the edge lifetime diverges with system size. Finite-size effects are therefore more significant, and the continued increase of the lifetime with $L$ suggests that the exponent is underestimated. At $K=-0.4$, on the other hand, $|K|$ is no longer small compared with $J$. The quantity $p|K|/J$ decreases to $0.8$, but the lifetimes are short and even decrease between $L=10$ and $L=12$, making the extracted slope only a rough estimate. Overall, our results confirm the qualitative dependence on $K$ predicted in Ref.~\cite{kemp2020}, but do not allow us to establish its asymptotic scaling quantitatively.

Finally, the lifetime cannot continue to grow indefinitely as $g$ decreases. For rational values of $K/J$, combinations of multiple spin flips can exactly compensate the energy $\Delta E_0$. These resonant processes, identified in Ref.~\cite{kemp2020}, also limit the order to which the zero mode can be constructed perturbatively~\cite{kemp2017}. At sufficiently small $g$, they determine the asymptotic power-law growth of the lifetime given in Eq.~\eqref{eq:KYL}. In our accessible range, however, the local exponent continues to increase as $g$ decreases, indicating that this asymptotic regime has not yet been reached. The exponential fit should therefore be understood as an effective description of the numerical data over $2\lesssim J/g\lesssim3.6$, rather than as the ultimate scaling at small $g$.
\subsection{Finite temperature}
\label{sec:temp}

An important question is how the lifetime of the edge memory changes with temperature. At finite temperature, an additional feature becomes relevant: the sign of the next-nearest-neighbor coupling $K$ can affect the edge dynamics. At infinite temperature, the two signs are equivalent. This follows from the exact mapping in Eq.~\eqref{eq:U}, under which the unitary $U$ transforms the Hamiltonian according to
\[
UH(g,K)U^\dagger=-H(g,-K),
\]
while leaving the edge spin unchanged. Since all states are equally populated at infinite temperature, reversing the sign of the Hamiltonian does not affect the dynamics of the edge memory. The two signs of $K$ therefore give identical results. At finite temperature, however, the sign of the Hamiltonian changes the thermal populations. The same mapping implies that the chain with coupling $-K$ at inverse temperature $\beta$ is equivalent to the chain with coupling $K$ at inverse temperature $-\beta$. Thus, cooling the frustrated chain ($K>0$) is equivalent to heating the unfrustrated chain ($K<0$) to negative temperature. The two signs of $K$ can consequently exhibit very different temperature dependences. In particular, both cases can be studied using a single calculation at $K=-0.3$, provided that both positive and negative values of $\beta$ are included.

We characterize the finite-temperature dynamics using the thermal autocorrelation function
\[
\Aedge(t)=\mathrm{Tr}[\rho_\beta Z_0(t)Z_0],
\qquad
\rho_\beta=\frac{e^{-\beta H}}{\mathrm{Tr}\,e^{-\beta H}},
\]
where $\beta$ is the inverse temperature. The thermal zero-mode weight $q_\beta$ is obtained from the same spectral decomposition as at infinite temperature, with each eigenstate weighted by its Boltzmann probability. We define the corresponding edge lifetime $\tedge(\beta)$ through
\[
\Aedge(\tedge)=q_\beta/2.
\]
We focus on the edge memory rather than the thermal OTOC, since for the system sizes accessible to us, the latter is strongly affected by finite-size fluctuations at low temperatures. The edge memory therefore provides a more reliable measure of how temperature affects the slow boundary dynamics.

\begin{figure}[t]
\includegraphics[width=\columnwidth]{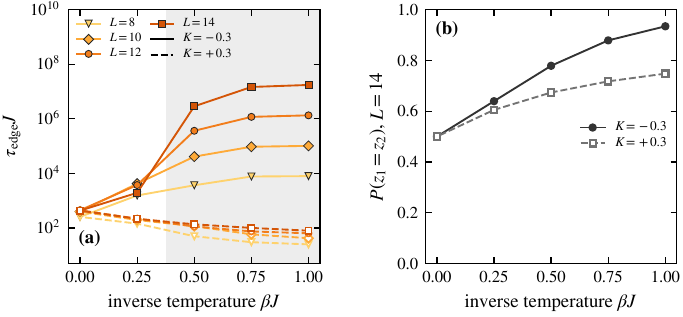}
\caption{Finite temperature at $g=0.4$. (a) Edge lifetime $\tedge$ as a function of inverse temperature for $K=-0.3$ (solid lines, filled symbols) and, via the exact mapping, $K=+0.3$ (dashed lines, open symbols), for $L=8$--14. In the shaded region the lifetimes of the unfrustrated chain are not converged in $L$. (b) Thermal probability $P(z_1=z_2)$ that the nearest ($j=1$) and next-nearest ($j=2$) neighbors of the edge spin are aligned, for $K=-0.3$ (filled symbols) and $K=+0.3$ (open symbols), for $L=14$.}
\label{fig:temp}
\end{figure}

The main result is that cooling has opposite effects for the two signs of $K$. For the unfrustrated chain with $K=-0.3$, the edge lifetime increases substantially as the temperature is lowered [solid lines in Fig.~\ref{fig:temp}(a)]. At infinite temperature, $\tedge\simeq436$ for $L=14$. Already at $\beta J=0.25$, it increases to values between $1.6\times10^3$ and $4.2\times10^3$, depending on the system size ($8\leq L\leq14$). For $\beta J\geq0.5$, the lifetime exceeds $4\times10^4$ for $L\geq10$ and continues to increase with system size. These values have not converged with $L$, and we therefore do not quote definite lifetimes in this regime [shaded region in Fig.~\ref{fig:temp}(a)]. The zero-mode weight increases simultaneously, from $q_\beta=0.74$ at $\beta=0$ to $0.90$ at $\beta J=1$. Cooling thus increases the fraction of the edge spin protected by the slow mode and further suppresses its decay.

The behavior is reversed for the frustrated chain with $K=+0.3$ [dashed lines in Fig.~\ref{fig:temp}(a)]. Here, cooling reduces the edge lifetime, which reaches $\tedge\approx80$ at $\beta J=1$ for $L=14$. The zero-mode weight also decreases, reaching $q_\beta\simeq0.53$. Although this lifetime is not fully converged either, the finite-size dependence is clear: it increases by approximately a factor of three between $L=8$, where $\tedge\approx25$, and $L=14$. Nevertheless, it remains far below the infinite-temperature value $\tedge\simeq436$. The reduction of the lifetime upon cooling is therefore robust, despite the remaining finite-size effects.

The exact mapping between the two signs of $K$ explains why their temperature dependences are opposite near infinite temperature. Since the frustrated chain at inverse temperature $\beta$ is equivalent to the unfrustrated chain at $-\beta$, their lifetimes are related by reflection about $\beta=0$. Consequently, their slopes at infinite temperature must have equal magnitudes and opposite signs. Thus, if cooling increases the lifetime for one sign of $K$, it must decrease it for the other. This opposite response is an exact consequence of the mapping to linear order in $\beta$ near infinite temperature. Its persistence up to $\beta J=1$, well beyond the linear regime, is established by our numerical results.

The physical origin of this behavior can be understood from the local energy costs associated with flipping the edge spin. The dominant leakage process is expected to involve the lower-energy channel, in which the flip costs $2(J-|K|)$. Whether this channel is available depends on the relative orientations of the nearest and next-nearest neighbors of the edge spin. For $K<0$, the lower-energy channel requires these two neighboring spins to point in opposite directions. Cooling the unfrustrated chain favors configurations in which they are aligned. As shown in Fig.~\ref{fig:temp}(b), the probability $P(z_1=z_2)$ increases from $1/2$ at infinite temperature to $0.93$ at $\beta J=1$. The lower-energy channel is therefore suppressed, leaving the edge spin to relax predominantly through the more expensive channel. This accounts for the increase in its lifetime.

For $K>0$, the situation is reversed. The lower-energy channel now requires the two neighboring spins to be aligned. Cooling the frustrated chain increases their alignment probability to $P(z_1=z_2)\simeq0.75$ at $\beta J=1$, making the lower-energy channel more accessible and thereby shortening the edge lifetime. Temperature therefore affects the edge memory not simply through the overall density of thermal excitations, but also by changing the local spin configurations that determine the energy cost of releasing information from the boundary.

Our finite-temperature results should be interpreted qualitatively, particularly at low temperatures. For the unfrustrated chain, the rapid growth of the lifetime with system size prevents a reliable finite-size extrapolation using exact methods. The frustrated chain also exhibits appreciable finite-size effects. A quantitative description of this regime will require tensor-network methods, such as those used in the recent study of a related interacting Majorana chain~\cite{tausendpfund2025}, which found an edge lifetime that grows exponentially with inverse temperature. Nevertheless, our results establish a clear reversal of the temperature dependence between the two signs of $K$ and identify the local edge energetics as its physical origin.

\section{Conclusions}
\label{sec:conclusions}

We have studied how information stored in the boundary spin of a nonintegrable Ising chain in its ordered phase is transferred to the bulk and scrambled there. Our central finding is that this process is bottlenecked by the boundary. Because flipping the edge spin is off-resonant, a large part of its weight is carried by an almost strong zero mode, which retains the edge information for very long times. This information leaks into the bulk only slowly, as the zero mode decays, and once it has entered the bulk it is scrambled quickly and almost uniformly. As a consequence, the edge-to-bulk OTOC does not develop a light cone. Apart from a weak front carried by the fast component of the edge spin, it grows slowly and nearly uniformly across the chain, at a rate set by the release of information from the edge rather than by the butterfly velocity of the bulk.

This picture becomes quantitative through the time-doubling law, $\Cbar(t)\simeq1-\Aedge(2t)$, which relates the bulk OTOC to the edge memory without any adjustable parameters. The factor of two arises because the OTOC measures the weight of the part of the operator that remains at the edge, that is, the square of its amplitude, which by time-reversal symmetry equals the memory at twice the time. In this sense, the law is the boundary counterpart of the slow OTOC tails caused by conserved quantities in the bulk~\cite{khemani2018,rakovszky2018,balachandran2021,balachandran2023}, with the slowly decaying zero mode playing the role of the conserved charge. The same mechanism also explains which operators are protected. Only operators that overlap with the zero mode inherit its long lifetime, whereas the transverse operators $X_0$ and $Y_0$, although they act on the same site, scramble as fast as bulk operators.

Since the time-doubling law ties scrambling to the edge memory, the lifetime of the zero mode directly sets the timescale of edge-to-bulk scrambling. In the range of fields accessible to us, this lifetime grows effectively exponentially in $J/g$. Its dependence on the next-nearest-neighbor coupling follows the trend predicted by Kemp, Yao, and Laumann~\cite{kemp2020}, although our window lies in the crossover regime and does not reach their asymptotic power law. Finally, the same local energetics that controls the lifetime also governs its dependence on temperature. Cooling closes or opens the cheapest leakage channel depending on the sign of $K$, and therefore lengthens or shortens the lifetime. An exact mapping between the two signs of $K$ guarantees this reversal near infinite temperature, and our numerics show that it persists to lower temperatures.

These predictions can be tested on current quantum simulators, where Ising chains with tunable transverse fields are realized with trapped ions and Rydberg atoms~\cite{zhang2017,bernien2017}, OTOCs have been measured~\cite{garttner2017,landsman2019,joshi2020,mi2021,braumuller2022}, and long-lived prethermal edge modes have been observed~\cite{jin2025}. The time-doubling law suggests a simple test: measure the edge autocorrelation function, which requires only random product states and local readout, and compare it with the edge-to-bulk OTOC at half the time. The most accessible regime is $g\approx0.45$--$0.5$, where the edge lifetime of about $10^2\,J^{-1}$ corresponds to a few microseconds for typical couplings, comparable to current coherence times. Beyond this model, the time-doubling law should also apply to other edge modes, such as Floquet $\pi$ modes~\cite{yates2019,yates2022} and interface zero modes~\cite{olund2025}, and its predicted breakdown when time-reversal symmetry is broken provides a direct test of its origin. Krylov-space and tensor-network methods~\cite{yates2020,yates2020dyn,tausendpfund2025} could further determine the asymptotic lifetime and the low-temperature regime. More broadly, our results suggest that simple autocorrelation measurements can serve as a proxy for scrambling in systems where OTOCs are hard to access.

\paragraph*{Data availability.} The data supporting the findings of this study are openly available at \cite{github_data_FTN}.

\appendix

\section{Derivation of the time-doubling law}
\label{app:derivation}

Throughout this appendix, overlaps between operators are written as infinite-temperature expectation values, $\mathrm{Tr}(A^\dagger B)/D$. Every operator can be expanded in Pauli strings $S$, that is, products of $I$, $X$, $Y$, and $Z$ on the individual sites. Pauli strings are Hermitian, square to one, and are orthonormal in the sense $\mathrm{Tr}(SS')/D=\delta_{SS'}$. A Hermitian operator has real coefficients in this basis. We write the evolved edge spin as
\begin{equation}
Z_0(t)=\sum_S c_S(t)\,S ,\qquad \sum_S c_S(t)^2=\frac{1}{D}\,\mathrm{Tr}\big[Z_0(t)^2\big]=1 ,
\label{eq:norm}
\end{equation}
where the second relation expresses that unitary time evolution preserves the total weight of the operator. The coefficient $c_S(t)^2$ is the weight of the string $S$ in $Z_0(t)$, and these weights always add up to one.

We first derive an exact identity for the memory at time $2t$. By definition, $\Aedge(2t)=\mathrm{Tr}[Z_0(2t)Z_0]/D=\mathrm{Tr}[e^{2iHt}Z_0e^{-2iHt}Z_0]/D$. Splitting each factor $e^{\pm2iHt}$ into two halves and using the cyclic property of the trace gives
\begin{align*}
\Aedge(2t)&=\frac{1}{D}\,\mathrm{Tr}\big[e^{iHt}Z_0e^{-iHt}\,e^{-iHt}Z_0e^{iHt}\big]\\
&=\frac{1}{D}\,\mathrm{Tr}\big[Z_0(t)\,Z_0(-t)\big].
\end{align*}
The memory at time $2t$ is therefore the overlap between the operator evolved forward by $t$ and the operator evolved backward by $t$. Next, we use time-reversal symmetry. The Hamiltonian of Eq.~\eqref{eq:H} and the operator $Z_0$ are real matrices in the $Z$ basis. Taking the complex conjugate of $Z_0(t)=e^{iHt}Z_0e^{-iHt}$ therefore only reverses the sign of $i$ in the exponents, so that $Z_0(t)^*=e^{-iHt}Z_0e^{iHt}=Z_0(-t)$. Backward evolution is thus the same as complex conjugation. Among the Pauli matrices, $X$ and $Z$ are real, whereas $Y$ is purely imaginary, $Y^*=-Y$. A Pauli string therefore satisfies $S^*=(-1)^{n_Y(S)}S$, where $n_Y(S)$ is the number of $Y$ operators in $S$. Since the coefficients $c_S$ are real, $Z_0(-t)=Z_0(t)^*=\sum_S(-1)^{n_Y(S)}c_S(t)\,S$. Inserting both expansions into the trace and using the orthonormality of the strings, all cross terms vanish, and we obtain Eq.~\eqref{eq:TR},
\begin{equation}
\Aedge(2t)=\sum_S(-1)^{n_Y(S)}c_S(t)^2=1-2\!\!\sum_{n_Y(S)\,\mathrm{odd}}\!\!c_S(t)^2 .
\label{eq:A1}
\end{equation}
The second form follows from the normalization in Eq.~\eqref{eq:norm}. This identity is exact and involves no approximation. It states that the memory at time $2t$ measures the imbalance between the weight of $Z_0(t)$ on strings with an even number of $Y$ operators and on strings with an odd number.

We now separate the slow edge part of the operator from the part that leaks into the bulk. At $t=0$ we decompose the edge spin as
\begin{equation}
Z_0=\sum_k\sqrt{q_k}\,\Psi_k+R,
\label{eq:dec}
\end{equation}
where the $\Psi_k$ are orthonormal, almost conserved operators localized within a few sites of the edge, and $R$ contains only spectral weight at frequencies of order $J$. A single almost strong zero mode corresponds to a single term in the sum, with $q_1=q$. We allow for several terms because the zero-frequency peak need not have a simple shape; for example, the two leakage channels discussed in Sec.~\ref{sec:tau} decay at different rates. A key property of the $\Psi_k$ is that they are real matrices. Because $H$ is real, its eigenvectors can be chosen real, and the matrix elements $\langle m|Z_0|n\rangle$ are then also real. The spectral filtering of Eq.~\eqref{eq:Psi}, which keeps only some of these real matrix elements, therefore produces a real matrix. A real Hermitian operator $O$ satisfies $O^*=O$, which by $S^*=(-1)^{n_Y(S)}S$ means that it contains only strings with an even number of $Y$ operators. The slow edge operators $\Psi_k$ are thus built entirely from even-$n_Y$ strings.

The fast part $R$ behaves like an ordinary bulk operator: it leaves the edge region within a short transient time $t_0$, which is between a few and about ten $J^{-1}$ in our data. For times $t\gg t_0$, we therefore split the evolved operator into two parts,
\begin{equation}
Z_0(t)=\mathcal{E}(t)+\mathcal{B}(t),
\label{eq:Z0t}
\end{equation}
where $\mathcal{E}(t)$ is the projection of $Z_0(t)$ onto operators localized near the edge, and $\mathcal{B}(t)$ is the remainder, which has spread into the bulk. Because $\mathcal{E}$ is a projection, the two parts are orthogonal, $\mathrm{Tr}(\mathcal{E}\mathcal{B})=0$, and their weights add up to one, $\mathrm{Tr}(\mathcal{E}^2)/D+\mathrm{Tr}(\mathcal{B}^2)/D=1$. On the slow timescale, the only operators that remain near the edge are the almost conserved $\Psi_k$, so $\mathcal{E}(t)$ lies in their span and contains only even-$n_Y$ strings. Inserting the decomposition into Eq.~\eqref{eq:A1}, the edge part contributes its full weight with a positive sign, whereas the bulk part contributes with signs that depend on $n_Y$,
\begin{equation}
\Aedge(2t)=\frac{1}{D}\,\mathrm{Tr}\big(\mathcal{E}^2\big)+\frac{1}{D}\,\mathrm{Tr}\big(\mathcal{B}^*\mathcal{B}\big),
\label{eq:A3}
\end{equation}
where $\mathrm{Tr}(\mathcal{B}^*\mathcal{B})/D=\sum_S(-1)^{n_Y(S)}b_S^2$, with $b_S$ the coefficients of $\mathcal{B}$. The second term contains the main approximation of the derivation. Once the leaked part has been scrambled by the chaotic bulk, its weight is spread over a very large number of strings, and strings with even and odd $n_Y$ are equally likely. Their contributions then cancel, $\mathrm{Tr}(\mathcal{B}^*\mathcal{B})/D\simeq0$, and Eq.~\eqref{eq:A3} reduces to Eq.~\eqref{eq:Enorm},
\begin{equation*}
\frac{1}{D}\,\mathrm{Tr}\big(\mathcal{E}^2\big)\simeq\Aedge(2t),\qquad \frac{1}{D}\,\mathrm{Tr}\big(\mathcal{B}^2\big)\simeq1-\Aedge(2t).
\end{equation*}
The weight that remains at the edge at time $t$ is therefore equal to the memory at time $2t$, and the weight that has leaked into the bulk is $1-\Aedge(2t)$. This result requires no assumption about the shape of the decay of $\Aedge$.

It is instructive to see how this works in the simplest case, in which each slow component decays exponentially. Suppose that
\begin{equation}
\Psi_k(t)=e^{-\Gamma_kt}\Psi_k+\mathcal{B}_k(t),
\label{eq:psidecay}
\end{equation}
where $\Gamma_k$ is the decay rate of the $k$th component and $\mathcal{B}_k(t)$ is the part that has leaked into the bulk, which is orthogonal to all operators localized near the edge. The memory is the overlap of $Z_0(t)$ with $Z_0$. The leaked operators $\mathcal{B}_k(t)$ and the fast part $R(t)$ have negligible overlap with the edge spin, so only the surviving amplitudes contribute, $\Aedge(t)\simeq\sum_kq_ke^{-\Gamma_kt}$. The edge part of the evolved operator is $\mathcal{E}(t)=\sum_k\sqrt{q_k}\,e^{-\Gamma_kt}\Psi_k$, and its weight is $\mathrm{Tr}(\mathcal{E}^2)/D=\sum_kq_ke^{-2\Gamma_kt}=\Aedge(2t)$, in agreement with the general result. This example shows the origin of the factor of two directly: the memory decays with the amplitude, at rate $\Gamma_k$, whereas the weight is the square of the amplitude and decays at rate $2\Gamma_k$.

We now relate the leaked weight to the OTOC. Consider a probe $V_j$, a single-site Pauli operator at site $j$, located farther from the edge than the localization length of the $\Psi_k$. The edge part $\mathcal{E}(t)$ does not reach site $j$ and therefore commutes with the probe, $[\mathcal{E}(t),V_j]\simeq0$. Only the leaked part contributes to the commutator, and the OTOC of Eq.~\eqref{eq:C} becomes
\begin{equation}
C_{0\to j}(t)\simeq\frac{1}{2D}\,\mathrm{Tr}\big([\mathcal{B}(t),V_j]^\dagger[\mathcal{B}(t),V_j]\big).
\label{eq:OTOCB}
\end{equation}
To evaluate this, we expand $\mathcal{B}=\sum_Sb_SS$. Each string either commutes with $V_j$, in which case $[S,V_j]=0$, or anticommutes with it, in which case $[S,V_j]=2SV_j$. Only anticommuting strings contribute, each with weight $4b_S^2$, and the prefactor $1/2D$ turns Eq.~\eqref{eq:OTOCB} into $2\sum_{\{S,V_j\}=0}b_S^2$, that is, twice the weight of $\mathcal{B}$ on strings that anticommute with the probe. In a scrambled operator, the four single-site operators $I$, $X$, $Y$, and $Z$ appear at site $j$ with equal weight. Two of them anticommute with $V_j$; for $V_j=Z_j$, these are $X_j$ and $Y_j$. Half of the weight of $\mathcal{B}$ therefore lies on anticommuting strings, and the OTOC equals the total leaked weight,
\begin{equation*}
C_{0\to j}(t)\simeq\frac{1}{D}\,\mathrm{Tr}\big(\mathcal{B}^2\big)\simeq1-\Aedge(2t).
\end{equation*}
This holds for every bulk probe, so the bulk average obeys the same relation, which is Eq.~\eqref{eq:doubling}. The argument uses only that the probe is a single-site Pauli operator, so it applies equally to $Z_j$ and $X_j$ probes. Applied to $\hat\Psi$ as the source instead of $Z_0$, the same steps give $\overline{C}_{\hat\Psi\to j}(t)\simeq1-A_{\hat\Psi}(2t)$.

The bounds of Eq.~\eqref{eq:band} follow when the slow part of the memory is a sum of exponentials, $\Aedge(t)\simeq\sum_kq_ke^{-\Gamma_kt}$ with $q=\sum_kq_k\le1$. For the lower bound, we note that all rates are nonnegative, so each term satisfies $e^{-2\Gamma_kt}\le e^{-\Gamma_kt}$. Hence $\Aedge(2t)\le\Aedge(t)$, and the time-doubling law gives $\Cbar(t)\ge1-\Aedge(t)$: the OTOC is always at least as large as the lost memory. For the upper bound, we use the Cauchy--Schwarz inequality with the vectors $\sqrt{q_k}$ and $\sqrt{q_k}\,e^{-\Gamma_kt}$,
\begin{equation}
\Big(\sum_kq_ke^{-\Gamma_kt}\Big)^2\le\Big(\sum_kq_k\Big)\Big(\sum_kq_ke^{-2\Gamma_kt}\Big),
\label{eq:CS}
\end{equation}
which reads $\Aedge(t)^2\le q\,\Aedge(2t)$. Hence $\Cbar(t)\le1-\Aedge(t)^2/q$, and since $q\le1$, also $\Cbar(t)\le1-\Aedge(t)^2$. The Cauchy--Schwarz inequality becomes an equality only when the two vectors are parallel, that is, when all components decay at the same rate. The upper bound $1-\Aedge(t)^2/q$ is therefore reached only for a single exponential decay.

Finally, the derivation shows when the time-doubling law should fail. First, it requires that the leaked weight has reached the probe, so it does not apply before the transit time from the edge to site $j$. Second, it requires $[\mathcal{E},V_j]\simeq0$, so it does not apply to probes within the localization length of the zero mode. Third, it requires that the leaked part is fully scrambled. If leakage is not slow compared with bulk scrambling, part of the leaked weight is still close to the edge and not yet spread uniformly over all Pauli operators, so less than half of it anticommutes with a distant probe. This makes $\Cbar$ smaller than $1-\Aedge(2t)$, consistent with the small negative deviations we observe at $g=0.3$, where bulk scrambling is slowest. Fourth, the exact identity of Eq.~\eqref{eq:A1} relies on time-reversal symmetry. If this symmetry is broken, for example by a field with both $X$ and $Y$ components, backward evolution is no longer equivalent to complex conjugation, and the law must be modified.

\section{Numerical methods and accuracy}
\label{app:methods}

All results in this paper are obtained with three numerical methods, chosen according to the system size and the quantity computed. For chains with up to $L=14$ sites we use exact diagonalization (ED), which gives the edge memory and the zero-mode weight $q(\omega_c)$ exactly, at all times. For OTOCs, exact evaluation is too expensive beyond $L=10$, so we use dynamical typicality, in which the trace over all states is replaced by an average over a few random states. For the largest chains, $L=16$ and 18, even diagonalization is too expensive, and we combine dynamical typicality with a Chebyshev expansion of the time-evolution operator, which only requires multiplying vectors by the sparse Hamiltonian. These methods limit us to $L\le18$. For the largest sizes, the OTOCs reach times $t\le150$ ($L=16$) and $t\le80$ ($L=18$), and the edge memory reaches $t\le3\times10^3$ ($L=16$) and $t\le600$ ($L=18$). Below we describe each method and how we checked its accuracy.

Exact diagonalization becomes much cheaper if we use the spin-flip parity $P=\prod_jX_j$. Because $P$ commutes with $H$, the Hamiltonian does not connect states of opposite parity and splits into two independent blocks, each of half the size. We construct these blocks in the basis $|s,\pm\rangle=(|s\rangle\pm|\bar s\rangle)/\sqrt2$, where $\bar s$ is the configuration obtained by flipping all spins of $s$, and $s$ runs over the configurations with $z_0=+1$. The edge spin $Z_0$ flips the parity, mapping $|s,+\rangle$ to $|s,-\rangle$. Its matrix elements therefore connect only the two blocks, $M_{mn}=\langle m,+|Z_0|n,-\rangle$, where $|m,+\rangle$ and $|n,-\rangle$ are the eigenstates of the two blocks. From the energies and these matrix elements, the edge memory $\Aedge(t)$ and the zero-mode weight $q(\omega_c)$ of Eq.~\eqref{eq:qwc} follow exactly and for arbitrary times, without any time stepping. As a check, the results agree with full ED, without the parity reduction, to $10^{-12}$ for $L=10$. The thermal quantities of Sec.~\ref{sec:temp} are obtained in the same way, by weighting each eigenstate with its Boltzmann factor. For the level statistics of Fig.~\ref{fig:levels}, we additionally use the reflection symmetry $j\to L-1-j$, which splits each parity block into two further blocks. Level repulsion can only be seen within a single symmetry sector, so this full resolution is necessary. We keep only the central part of each spectrum, leaving out the lowest and highest levels, where the density of states is low, and average $\langle r\rangle$ over the two reflection sectors, weighted by their dimensions.

The OTOC requires a trace over all $D=2^L$ states, which is expensive. Dynamical typicality~\cite{elsayed2013,xu2024tutorial} avoids this: for a Haar-random state $|\psi\rangle$, the expectation value $\langle\psi|O|\psi\rangle$ equals the infinite-temperature average $\mathrm{Tr}(O)/D$ up to a statistical error of order $D^{-1/2}$, which becomes small in large systems. For $L=12$, we therefore compute $C_{W\to V}(t)=\frac12\big\|[W(t),V]\,|\psi\rangle\big\|^2$ and average over two or three random states. We work in the energy eigenbasis, where the time evolution is simply a phase factor $e^{-iE_mt}$ for each eigenstate, so $W(t)$ is applied exactly at any time. To check the accuracy, we compared this method with the exact trace for $L=10$. The largest deviation over all times and probe sites is below $0.05$ with two random states, about twice the typical error $2^{-L/2}/\sqrt2\approx0.022$, as expected for a statistical error that scales as $D^{-1/2}$. For $L=12$, the statistical error is below $0.02$.

For $L=16$ and 18, the Hilbert space is too large for diagonalization. We therefore evolve the random states directly in time, using a Chebyshev expansion of $e^{-iHt}$~\cite{talezer1984,weisse2006}. This method requires only repeated multiplication of a vector by the sparse Hamiltonian, and its accuracy can be controlled precisely. We first rescale $H$ by a rigorous upper bound $a$ on its spectral radius, so that its spectrum lies between $-1$ and $1$. We then divide each time interval into steps $\delta t$ with $a|\delta t|\le40$, and choose the number of Chebyshev terms in each step large enough that the truncation error is below machine precision. The time evolution itself is therefore essentially exact, and the only remaining error is the statistical error of the random state. To check this, we computed the edge memory for $L=14$ with this method and a single random state. It deviates from the exact ED result by at most $0.011$ at all times, consistent with the expected typical error $2^{-L/2}\approx0.008$. For $L=16$ and 18, we use a single random state for each run. The statistical error, $2^{-L/2}$, is then $0.004$ for $L=16$ and $0.002$ for $L=18$, smaller than at $L=12$ because the error decreases with system size.
\bibliography{refs}

\end{document}